# Voltage and Energy-Delay Performance of Giant Spin Hall Effect Switching for Magnetic Memory and Logic

Sasikanth Manipatruni, Dmitri E. Nikonov, and Ian A. Young

*Exploratory Integrated Circuits, Intel Components Research, Intel Corp, Hillsboro, OR 97124*

In this letter, we show that Giant Spin Hall Effect (GSHE) MRAM can enable better energy-delay and voltage performance than traditional MTJ based spin torque devices at scaled nanomagnet dimensions (10-30 nm). Firstly, we derive the effect of dimensional scaling on spin injection efficiency, voltage-delay and energy-delay of spin torque switching using MTJs and GSHE and identify the optimum electrode geometry for low operating voltage (<0.1 V), high speed (>10 GHz) operation. We show that effective spin injection efficiency >100 % can be obtained using optimum spin hall electrode thickness for 30 nm nanomagnet widths. Finally, we derive the energy-delay trajectory of GSHE and MTJ devices to calculate the energy-delay product of GSHE and MTJ devices with an energy minimum at the characteristic time of the magnets. Optimized GSHE devices when combined with PMA can enable MRAM with scaled nanomagnets (30 nm X 60 nm), ultra-low voltage operation (< 0.1 V), fast switching times (10 ps) and switching energy as low as 100 aJ/bit.

**AUTHOR EMAIL ADDRESS:** sasikanth.manipatruni@intel.com

Direct transfer of spin angular momentum to manipulate nanomagnets using spin polarized currents has opened several avenues for applications in memory and logic applications [1-5]. In particular, spin transfer torque (STT) based magnetic RAM has emerged as a promising universal memory option due to the potential for high density and non-volatility. However, traditional STT devices are reliant on coherent Magnetic Tunnel Junctions (MTJs) formed by orbital symmetry dependent filtering using oxides (e.g MgO) [6]. MTJ based technology has the following fundamental drawbacks for compatibility with future scaled complementary metal oxide semiconductor (CMOS) technologies [7] a) Incompatibility of the high operating voltages required for large tunnel currents with scaled CMOS b) Large access transistor sizing for the large drive current requirement c) Reliability issues due to high voltage and current conduction through the MgO tunnel junctions. Hence it is of great interest to pursue alternate technologies for STT which provide high spin polarization at low voltage & low current operation.

Recently discovered Giant Spin Hall Effect (GSHE) [8-10], the generation of large spin currents transverse to the charge current direction in specific high-Z metals (such as Pt [11-13, 8-9], β-Ta [9], Beta-W [10], Doped Cu [e.g 14-15]) is a promising solution to the voltage, current scaling and reliability problems. Due to the relatively low resistivity of GSHE-metals compared to MTJs the write voltages compatible with future CMOS technology nodes can be expected while the required current density is reduced. However, the exact nature of the dimensional scaling of GSHE to 10 nm cell sizes is not clear.

In this letter, we derive the dimensional scaling trends for spin injection efficiency; write voltage, write energy & delay for GSHE based spin torque switching. We show that spin hall effect switching provides significant performance improvement in write energy compared to in-plane and perpendicular spin torque switching employing magnetic tunnel junctions. We also show that GSHE switching for scaled nanomagnets can allow ultra-low voltage operation.

A typical geometry of a 3-terminal memory cell with a spin Hall Effect induced write mechanism and MTJ based read-out is shown in figure 1. An example material stack [16] is shown in figure 1A, comprising of a free layer nanomagnet in direct contact with GSHE metal. A nominal MTJ stack comprising of free layer (FM1), MgO tunneling oxide, a fixed magnet (FM2) with Synthetic Anti-Ferro-magnet (SAF) and Anti-Ferromagnet (AFM) are shown. The SAF layer allows for cancelling the dipole fields around the free layer. A wide combination of materials has been studied for this material stacking. The write electrode consists of a GSHE metal made of β-Tantalum (β-Ta), β-Tungsten (β-W) or Pt [9, 10], the write electrode transitions into a normal high conductivity metal (Cu) to minimize the write electrode resistance. The top view of the device is shown in figure 1B where the magnet is oriented along the width of the GSHE electrode for appropriate spin injection. The magnetic cell is written by applying a charge current via the GSHE electrode. The direction of the magnetic writing is decided by the direction of the applied charge current. Positive currents (along +y) produce a spin injection current with transport direction (along +z) and spins pointing to (+x) direction. The injected spin current in-turn produces spin torque to align the magnet in the +x or –x direction. The transverse spin current ($\vec{I}_s = \vec{I}_\uparrow - \vec{I}_\downarrow$ with spin direction $\hat{\sigma}$) for a charge current ($\vec{I}_c$) in the write electrode is given by

$$\vec{I}_s = P_{she}(w, t, \lambda_{sf}, \theta_{SHE})(\hat{\sigma} \times \vec{I}_c) \qquad (1)$$

Where $P_{SHE} = (\vec{I}_\uparrow - \vec{I}_\downarrow)/(\vec{I}_\uparrow + \vec{I}_\downarrow)$ is the spin hall injection efficiency which is the ratio of magnitude of transverse spin current to lateral charge current, $w$ is the width of the magnet, $t$ is the thickness of the GSHE metal electrode, $\lambda_{sf}$ is the spin flip length in the GSHE metal, $\theta_{GSHE}$ is the spin hall angle for the GSHE-metal to FM1 interface. The injected spin angular momentum responsible for spin torque is given by $\vec{S}=\hbar\vec{I}_s/2e$.

We show that effective spin injection efficiency >100 % can be obtained using optimum spin hall electrode thickness for 30 nm nanomagnet widths. We show the spin hall injection efficiency (ratio of spin current injected to the charge current in the electrode) as a function of electrode thickness has an optimum value at 2-3 nm electrode thickness. The spin hall injection efficiency is given by:

$$P_{she} = \frac{I_{sz}}{I_{cy}} = \frac{\pi w}{4t}\theta_{SHE}\left(1-\sec h\left(\frac{t}{\lambda_{sf}}\right)\right) \qquad (2)$$

In fig. 2A we show the spin hall injection efficiency (spin polarization ratio) as a function of thickness of the spin hall metal electrode for 30 nm width ($w$) free layer. We assumed a 1.5 nm spin flip length for the GSHE metals with GSHE angles as follows; Pt (0.07 [9]), β-Tantalum (-0.15 [9]), β-Tungsten (0.3 [10]). We show that the spin hall injection efficiency (SHIE) may exceed 100 % for scaled nanomagnets for optimum GSHE electrode thickness. In fig. 2B we show the spin polarization ratio as a function of nanomagnet width. The SHIE decreases monotonically with scaled magnet dimensions. However, the SHIE remains >100 % for β-Tungsten even with highly scaled magnet dimensions. Hence, GSHE enhancement may be of utility in highly scaled magnetic memory applications. Higher spin injection ratio also implies lower required charge current with associated benefits in the control transistor areal footprint.

We show that the GSHE devices scale to operating voltages <0.2 V using an analytical scaling model for voltage and write delay of GSHE and MTJ memory devices. The analytical relationship connecting the switching time to the write voltage of a spin torque memory with critical voltage $v_c$ is given by:

$$\tau = \frac{\tau_0 \ln(\pi/2\theta_0)}{(v/v_c - 1)} \tag{3}$$

Where

$$v_{cSHE} = 8\rho I_c \left( \theta_{she} \left( 1 - \operatorname{sech}\left(\frac{t}{\lambda_{sf}}\right) \right) \pi L \right)^{-1} \tag{4}$$

Where $\theta_0 = \sqrt{k_B T / 2E_b}$ is the effect of stochastic variation due to thermal noise is, $E_b = \frac{1}{2}\mu_0 M_s H_k V$ is the thermal barrier of the magnet of volume V, saturation magnetization $M_s$, anisotropy $H_k$ and $\tau_0 = M_s V e / I_c P \mu_B$ is the characteristic time. $I_c$ is the critical current for spin torque induced magnetic switching. We have verified the validity of the equation 3 via stochastic spin torque simulations of a nanomagnet. We compare the switching time, voltage relationship of GSHE-device with the MTJ with nominal device parameters for a nanomagnet with 40 kT thermal barrier. We assumed in-plane magnets for this comparison. The assumed material properties are shown in figure 3. From figure 3A, we can see that the operating voltage range for MTJ devices is restricted to higher voltages than spin hall devices due to the high resistance of a magnetic tunnel junction device. Spin Hall devices comprised of any of the three considered metals will enable operating the write voltages as low as 100 mV. Secondly, GSHE does not suffer from the polarization degradation due voltage effects [17] which limit the peak injected currents and operating speed of MTJs to > 1 ns. The ability to operate in the 10 GHz switching frequency range using <0.1 V can have significant utility in spin logic devices [4, 5].

We derive the optimum electrode dimensions and relative switching energy of STT-MTJ devices using GSHE based STT writing of the MTJ devices (GSHE-MTJ) for identical nanomagnet dynamics. For identical switching dynamics (i.e identical delay and critical currents), the relative switching energy of GSHE writing to STT-MTJ devices is:

$$\frac{E_{she}}{E_{MTJP}} = \left(\frac{P_{MTJ}^2}{R_{MTJP}}\right)\left(\frac{R_{she}}{P_{she}^2}\right) = \left(\frac{P_{MTJ}^2}{R_{MTJP}}\right)\frac{32\rho_{she}t}{\pi^2 w l \theta_{SHE}^2 \left(1-\sec h\left(\frac{t}{\lambda_{sf}}\right)\right)^2} \quad (5)$$

To understand the effect of dimensional scaling, the ratio of the energy required to switch using identical nanomagnets (with identical barrier, damping & critical currents) are plotted in figure 4. We note that the relative energy scales with resistance of the write electrode and inversely proportional to the square of the spin injection efficiency. For MTJ devices, the product $R_{MTJ}/P_{MTJ}^2$ is fundamentally constrained since reducing the tunneling resistance is coupled with reducing spin polarization. However this trade off may be broken for GSHE devices if materials with higher $\theta_{SHE}$ & low resistivity ($\rho_{SHE}$) can be identified [15, 18-19], since the product $R_{SHE}/P_{SHE}^2$ is proportional to $\rho_{SHE}/\theta_{SHE}^2$

The relative energy of the GSHE as a function of thickness of the GSHE-electrode shows an optimum near $\lambda_{sf}$ - 3 $\lambda_{sf}$ with 10X-100X improvement in relative energy over STT based MTJs. This optima can be attributed to dimensional scaling terms in equation 5. We also note that the broad minima also allows for less sensitivity to line width, alignment and thickness variations invariably associated to large density memory/logic devices. Improvement in energy per switching state transition on the order of 10X to 100X can be expected with magnet sizes of 30 nm X 60 nm, as shown in figure 4A. The relative energy improvement is retained for highly scaled magnets with widths of 10-20 nm using optimum GSHE-electrode geometry. The relative energy advantage is significantly higher for larger nanomagnet widths and may have implications in spin hall based logic device design having high fan-out.

We now compare the energy-delay trajectory of GSHE and MTJ (GSHE-MTJ) devices for in-plane magnet switching as the applied write voltage is varied. The energy-delay trajectory (for in-plane switching) can be written as:

$$E(\tau) = R_{write} I_{co}^2 \frac{\left(\tau + \tau_o \ln\left(\frac{\pi}{2\theta_0}\right)\right)^2}{\tau} = \frac{4}{\hbar^2} \frac{R_{write}}{P^2} \frac{1}{\tau} \left(\mu_0 e \alpha M_s V \left(H_k + \frac{M_s}{2}\right)\left(\tau + \tau_o \ln\left(\frac{\pi}{2\theta_0}\right)\right)\right)^2 \quad (6)$$

Where $R_{write}$ is the write resistance of the device ($R_{GSHE}$ or $R_{MTJ-P}$, $R_{MTJ-AP}$), P is the spin current polarization ($P_{GSHE}$ or $P_{MTJ}$), $\mu_0$ is vacuum permeability, e is the electron charge. One can see that the energy at a given delay is directly proportional to the square of the gilbert damping. We also note that $\tau_0 = M_s V e / I_c P \mu_B$ varies as the spin polarization varies for various GSHE metal electrodes (figure 2). The combined effect of spin hall polarization, damping and resistivity of the spin hall electrodes is captured in equation 6 and plotted in figure 5A. All the cases considered in figure 5A assume a 30 X 60 nm magnet with 40 kT thermal energy barrier and 3.5 nm GSHE electrode thicknesses. The energy-delay trajectories of the devices are obtained assuming a voltage sweep from 0-0.7 V in accordance to voltage restrictions of scaled CMOS [6]. The energy-delay trajectory of the GSHE-MTJ devices exhibits broadly two operating regions A) Region 1 where the energy-delay product is approximately constant ($\tau_d < M_s V e / I_c P \mu_B$) B) Region 2 where the energy is proportional to the delay $\tau_d > M_s V e / I_c P \mu_B$. The two regions are separated by energy minima at $\tau_{opt} = M_s V e / I_c P \mu_B$ where minimum switching energy is obtained for the spin torque devices. The energy-delay trajectory of the STT-MTJ devices is limited with a minimum delay of 1 ns for in-plane devices at 0.7 V maximum applied voltage, the switching energy for P-AP and AP-P are in the range of 1 pJ/write. In contrast, the energy-delay trajectory of GSHE-MTJ (in-plane anisotropy) devices can enable switching times as low as 20 ps (β-W with 0.7 V, 20 fJ/bit) or switching energy as small as 2 fJ (β-W with 0.1 V, 1.5 ns switching time).

Finally, we compare the energy-delay of GSHE-MTJ and STT-MTJ devices for in-plane and perpendicular magnetic anisotropy (PMA). The spin orientation of the injected electrons using a planar GSHE electrode does not directly allow switching a PMA device (see Equation 1). However, alternative techniques may be feasible where a remnant dipole filed is applied [9], or an exchange bias filed is applied [20] or a Precessional spin torque switching is realized [21] such that a non-collinear spin torque can switch a PMA device. The energy-delay function of GSHE-MTJ and STT-MTJ switching with PMA is given by:

$$E(\tau) = R_{write} I_{co}^2 \frac{\left(\tau + \tau_o \ln\left(\frac{\pi}{2\theta_0}\right)\right)^2}{\tau} = \frac{4}{\hbar^2} \frac{R_{write}}{P^2} \frac{1}{\tau} \left(\mu_0 e \alpha M_s V H_k \left(\tau + \tau_o \ln\left(\frac{\pi}{2\theta_0}\right)\right)\right)^2 \quad (7)$$

The effect of PMA on the switching energy-delay of GSHE-MTJ and STT-MTJ based devices is shown in figure 6 B. Combining PMA with GSHE will provide significant improvements largely attributed to the lower critical currents of PMA due to the absence of demagnetizing fields which constrain the operation of the in-plane magnet dynamics. We see that combining GSHE and PMA for high speed magnetic switching can reduce the switching energy 3 orders of magnitude. From figure 5B, we can see that at 1 ns delay the switching energy of the PMA-GSHE-MTJ device compared to PMA-STT-MTJ is device is reduced to 100 aJ/bit from 100 fJ/bit. Alternatively, PMA-GSHE-MTJ devices can be used to reduce the switching time to 10 ps thus enabling ultra-fast memory and logic devices operating at low voltages. The PMA β-W design (red line with solid squares in figure 5B) corresponds to an energy-delay of 50 aJ.ns using an optimum geometry spin hall electrode ($t_{GSHE}$=2.3 $\lambda_{sf}$, $P_{GSHE}$ ~ 170 %) with 30 X 60 nm magnets with 40 kT thermal barrier representing a >100 X improvement in energy-delay while operating under a low voltage of < 0.1 V.

In conclusion, we describe the voltage, spin injection and energy-delay scaling of giant spin Hall Effect switching. We show that spin injection efficiency of spin hall effect switching can exceed 100 % even with highly scaled nanomagnets. We also identify the optimum electrode geometry for spin injection efficiency (to minimize the write current) and the optimum switching energy conditions. Using the voltage-delay scaling relation for GSHE devices we show that GSHE devices can enable high speed magnetic switching with future CMOS compatible voltages (0.1-0.4 V). Using an energy-delay scaling relation for GSHE-MTJ devices, we show that GSHE devices enable a much lower energy-delay product (PMA: 50-100 aJ.ns, InP: 0.4-3 fJ.ns) than STT-MTJ devices enabling 100 aJ/switching with switching times as low as 10 ps. Continued discovery and improvements in spin hall angle and conductivity of the spin hall metals may further enable the development of spintronic memory and logic. Hence, Giant Spin Hall Effect devices represent a significant opportunity for the development and integration of CMOS compatible spintronic memory and logic devices.

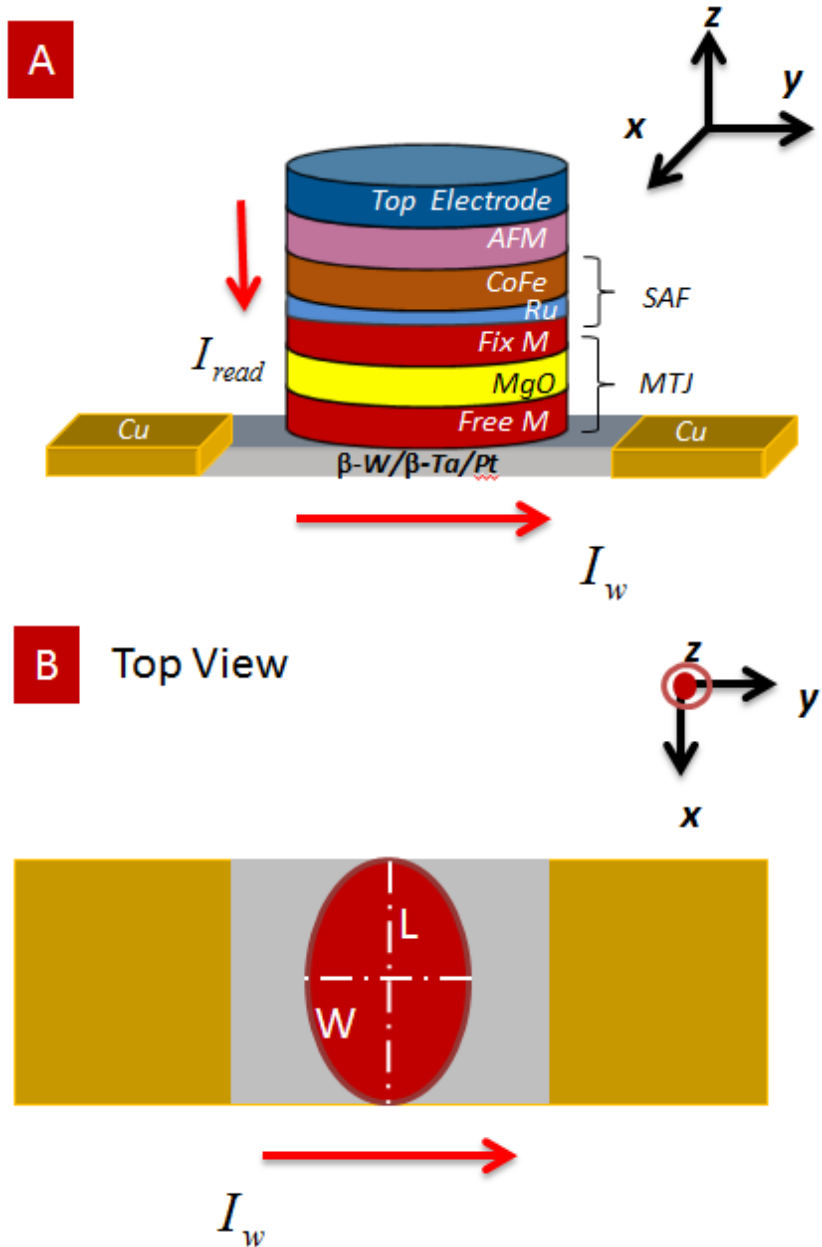

Figure 1 A) A Three Terminal Spin Hall Memory Device with spin hall effect write electrode and MTJ based readout B) Top view of the cell showing the orientation of the free layer magnet and the GSHE metal.

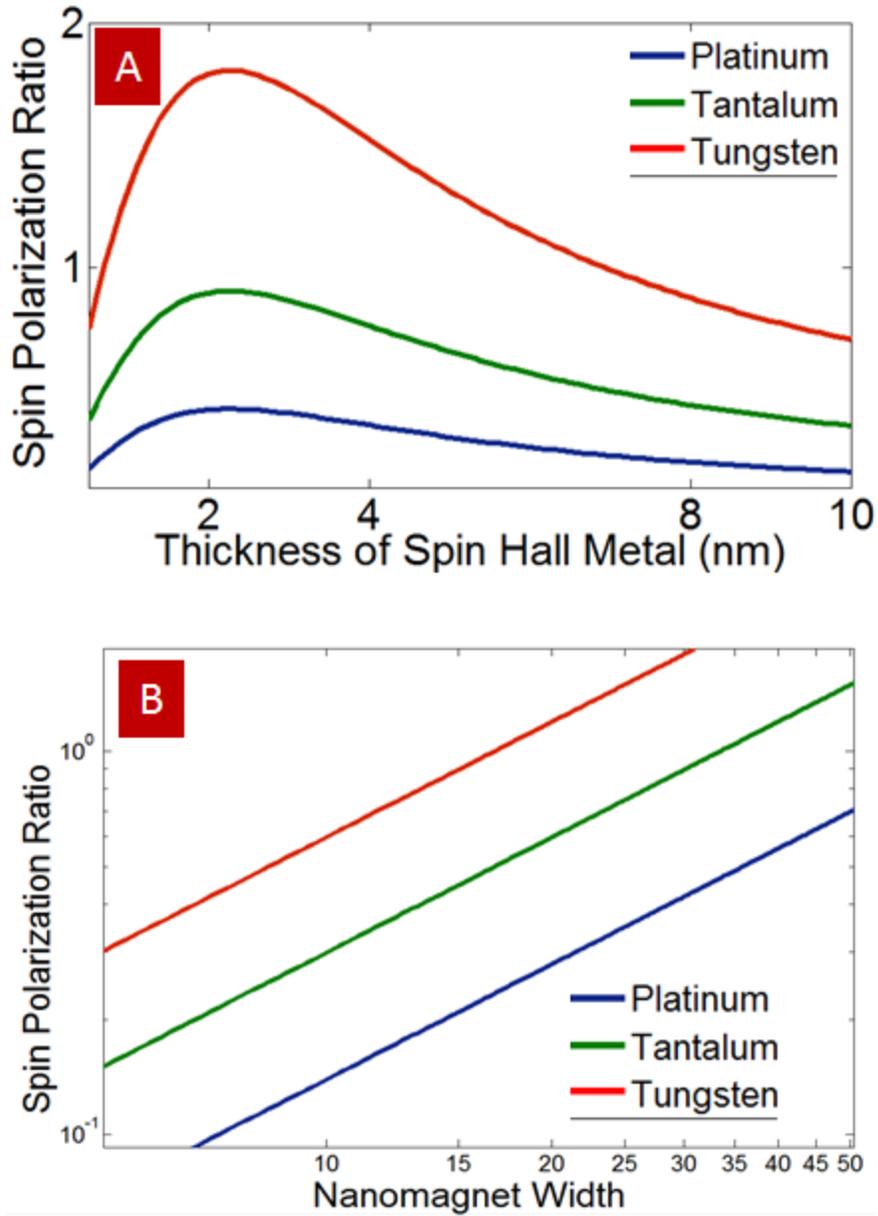

Figure 2 A) Effect Spin Hall Metal thickness on the spin polarization ratio for a nanomagnet width w= 30 nm (Fig 1B) B) Effect of nanomagnet width on the spin polarization ratio (for an electrode width of 4 nm)

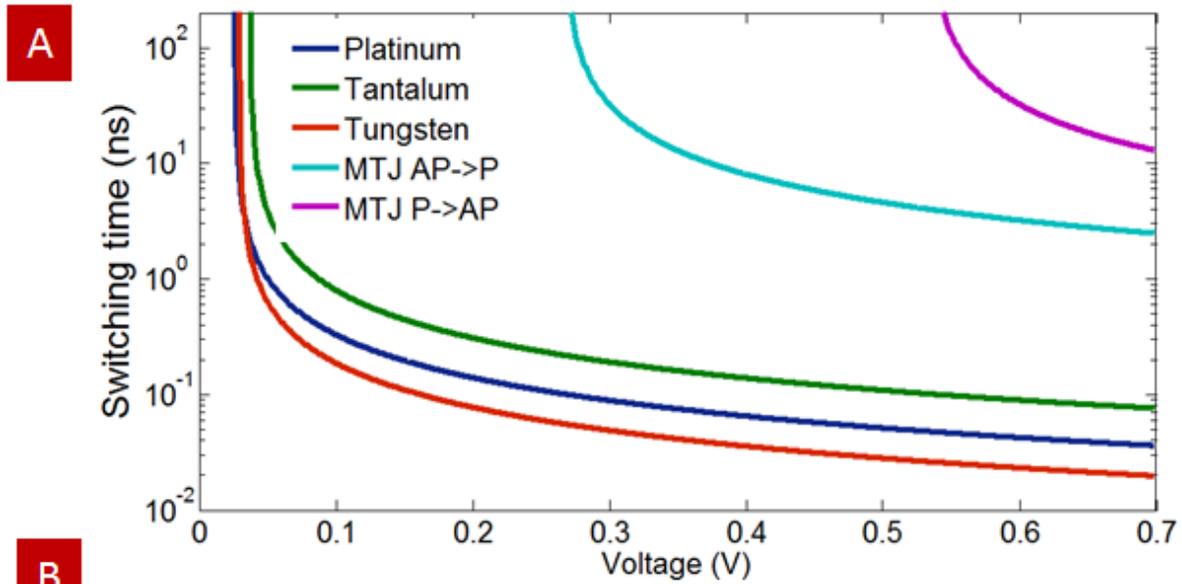

| Device | Resistivity (ρ) | Spin Hall Angle ( to CoFeB) θ$_{SHE}$ | Gilbert Damping (α) |
|---|---|---|---|
| Pt-SHE | 20 μΩ.cm | 0.07 | 0.025 |
| Ta-SHE | 190 μΩ.cm | -0.15 | 0.008 |
| W-SHE | 200 μΩ.cm | 0.3 | 0.012 |
| | RA Product | Spin Injection Efficiency | |
| MTJ | 5 Ω.μm² (P) 10 Ω.μm² (AP) | 80 % | 0.007 |

Figure 3 A) Switching time vs applied voltage to STT switching device using GSHE or MTJ based writing mechanism B) Assumed material and transport parameters.

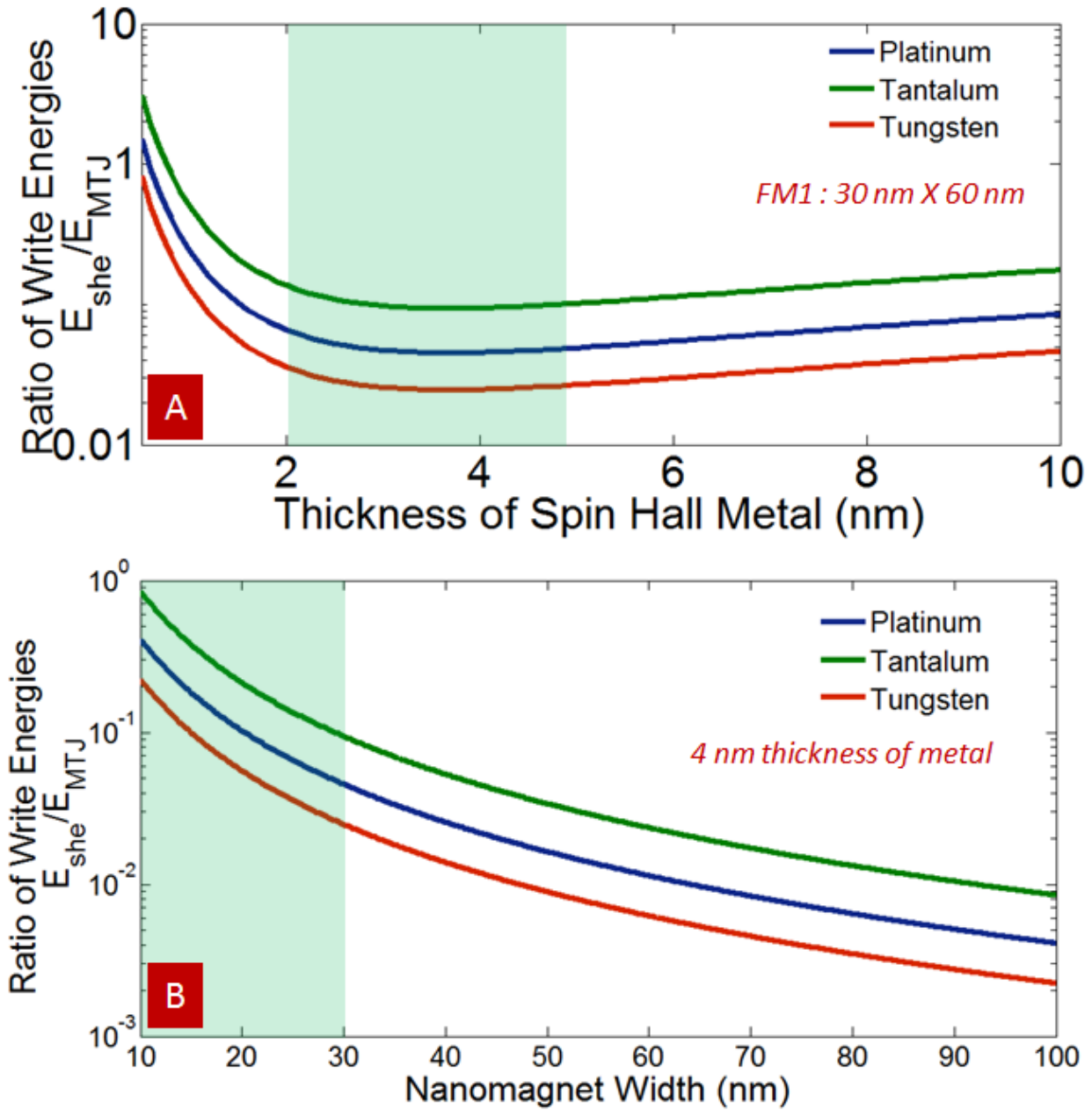

Figure 4: Relative switching energy for GSHE and MTJ based magnetic memory writing A) for varying spin hall metal electrode thickness for free layer magnet of 30 nm X 60 nm size B) for varying nanomagnet width for a GSHE metal with thickness 4 nm

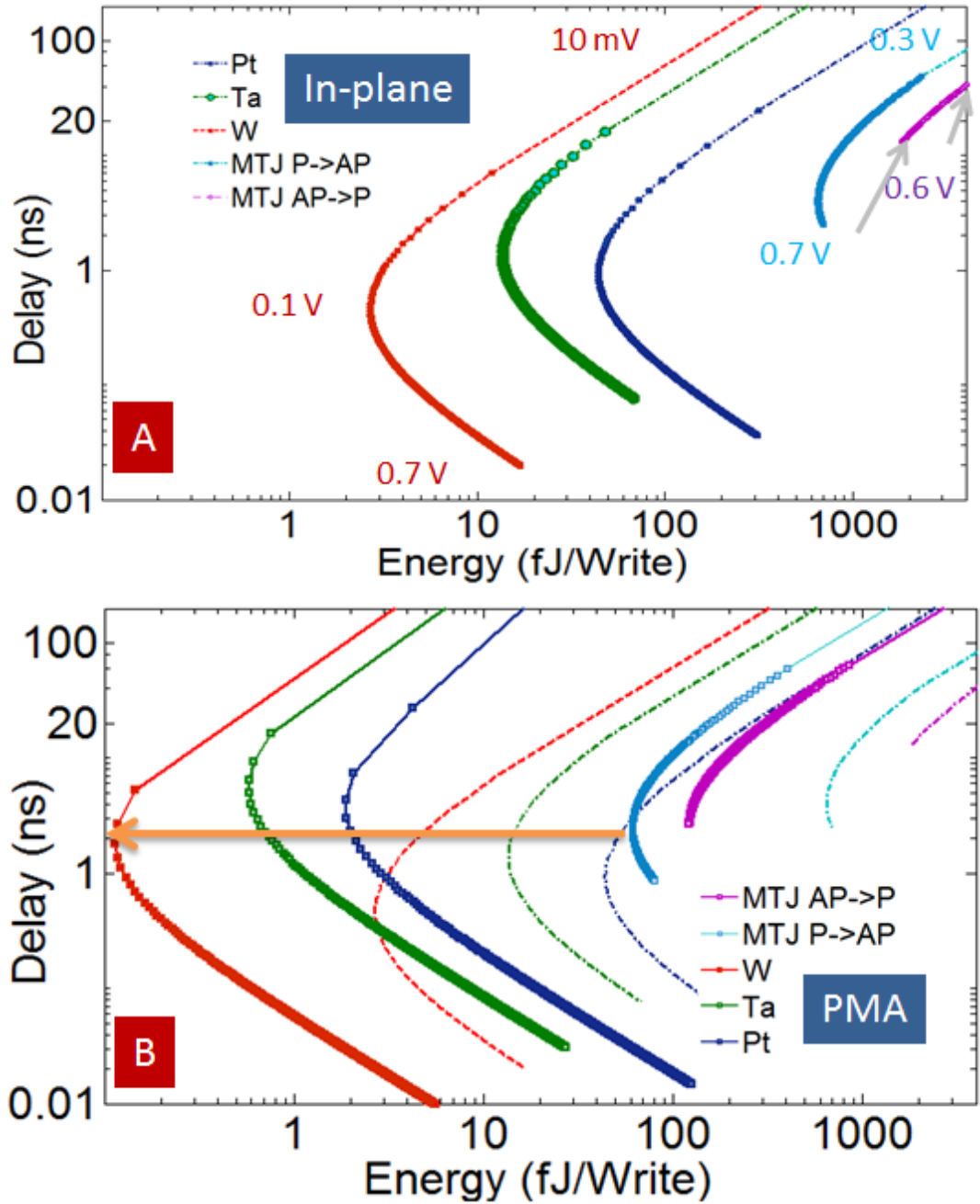

Figure 5: Energy-delay trajectory for GSHE switching with A) in-plane nanomagnet switching with write voltages B) PMA nanomagnet switching